\documentclass[preprint2]{aastex}

\usepackage{xcolor}
\usepackage{balance}

\slugcomment{Not to appear in Nonlearned J., 45.}

\shorttitle{Diffuse gamma-ray flux}
\shortauthors{KASCADE-Grande}

\begin{document}

\title{KASCADE-Grande limits on the isotropic diffuse gamma-ray flux between 100 TeV and 1 EeV}

\author{
W.D.~Apel$^{1}$, J.C.~Arteaga-Vel\'azquez$^{2}$, K.~Bekk$^{1}$, M.~Bertaina$^{3}$, J.~Bl\"umer$^{1,4}$-\footnote{now KIT, Division V - Physics and Mathematics}, H.~Bozdog$^{1}$, I.M.~Brancus$^{5}$, E.~Cantoni$^{3,6}$-\footnote{Now at: INRIM, Torino, Italy}, A.~Chiavassa$^{3}$, F.~Cossavella$^{4}$-\footnote{Now at: DLR Oberpfaffenhofen, Germany}, K.~Daumiller$^{1}$, V.~de Souza$^{7}$, F.~Di~Pierro$^{3}$, P.~Doll$^{1}$, R.~Engel$^{1}$, Z.~Feng$^{8}$, D.~Fuhrmann$^{9}$-\footnote{Now at: University of Duisburg-Essen, Germany}, A.~Gherghel-Lascu$^{5}$, H.J.~Gils$^{1}$, R.~Glasstetter$^{9}$, C.~Grupen$^{10}$, A.~Haungs$^{1}$, D.~Heck$^{1}$, J.R.~H\"orandel$^{11}$, T.~Huege$^{1}$, K.-H.~Kampert$^{9}$, D.~Kang$^{1}$, H.O.~Klages$^{1}$, K.~Link$^{4}$, P.~{\L}uczak$^{12}$, H.J.~Mathes$^{1}$, H.J.~Mayer$^{1}$, J.~Milke$^{1}$, B.~Mitrica$^{5}$, C.~Morello$^{6}$, J.~Oehlschl\"ager$^{1}$, S.~Ostapchenko$^{13}$, T.~Pierog$^{1}$, H.~Rebel$^{1}$, M.~Roth$^{1}$, H.~Schieler$^{1}$, S.~Schoo$^{1}$, F.G.~Schr\"oder$^{1}$, O.~Sima$^{14}$, G.~Toma$^{5}$, G.C.~Trinchero$^{6}$, H.~Ulrich$^{1}$, A.~Weindl$^{1}$, J.~Wochele$^{1}$, J.~Zabierowski$^{12}$ KASCADE-Grande Collaboration}

\affil{
$^1$ Institut f\"ur Kernphysik, KIT - Karlsruhe Institute of Technology, Germany\\
$^2$ Universidad Michoacana, Inst.~F\'{\i}sica y Matem\'aticas, Morelia, Mexico\\
$^3$ Dipartimento di Fisica, Universit\`a degli Studi di Torino, Italy\\
$^4$ Institut f\"ur Experimentelle Teilchenphysik, KIT - Karlsruhe Institute of Technology, Germany\\
$^5$ Horia Hulubei National Institute of Physics and Nuclear Engineering, Bucharest, Romania\\
$^6$ Osservatorio Astrofisico di Torino, INAF Torino, Italy\\
$^7$ Universidade de S$\tilde{a}$o Paulo, Instituto de F\'{\i}sica de S\~ao Carlos, Brasil\\
$^8$ Institute of High Energy Physics, Beijing, China\\
$^9$ Fachbereich Physik, Universit\"at Wuppertal, Germany\\
$^{10}$ Department of Physics, Siegen University, Germany\\
$^{11}$ Dept. of Astrophysics, Radboud University Nijmegen, The Netherlands\\
$^{12}$ National Centre for Nuclear Research, Department of Astrophysics, Lodz, Poland\\
$^{13}$ Frankfurt Institute for Advanced Studies (FIAS), Frankfurt am Main, Germany \\
$^{14}$ Department of Physics, University of Bucharest, Bucharest, Romania\\
}

\email{donghwa.kang@kit.edu, fengzy@ihep.ac.cn, andreas.haungs@kit.edu} \date{\today}

\begin{abstract}
KASCADE and KASCADE-Grande were multi-detector installations to measure 
individual air showers of cosmic rays at ultra-high energy. 
Based on data sets measured by KASCADE and KASCADE-Grande, 
90\% C.L. upper limits to the flux of $\gamma$ rays in the primary cosmic ray flux
are determined in an energy range of $10^{14} - 10^{18}$ eV. 
The analysis is performed by selecting air showers with a low muon content as expected for 
gamma-ray induced showers compared to air showers induced by energetic nuclei.
The best upper limit of the fraction of $\gamma$ rays to the 
total cosmic ray flux is 
obtained at $3.7\times10^{15}$ eV with $1.1\times10^{-5}$. Translated in an absolute 
$\gamma$-ray flux this sets constrains on some fundamental astrophysical models, 
such as the distance of sources for at least one of the IceCube neutrino excess models. 
\end{abstract}


\section{Introduction}
High-energy gamma rays represent a very small but also very important 
fraction of primary cosmic rays. Their importance derives
from the fact that they, other than the charged particles which 
constitute the bulk of cosmic ray primaries, are not deflected by 
interstellar magnetic fields. Hence, their direction of incidence 
on Earth points back to their origin.
Also for the case of
non-observation of gamma rays
in a certain energy range, 
upper limits on the flux provide important information.

In addition, investigations of the galactic and extragalactic 
diffuse gamma-ray emission are potentially able to give information 
about the source and propagation of galactic cosmic rays.
The measured flux of diffuse gamma-rays and its spectrum
thus would provide new insights into the acceleration of cosmic rays.
High-energy diffuse $\gamma$ rays are the sum of contributions from several components. 
One is the cascading products by the collision of cosmic rays 
with interstellar gas and dust in the disk of the Galaxy (Berezinsky et al. 1993).
In this case, the predicted integral intensity is concentrated in the galactic plane.
The other one is due to electromagnetic cascades induced by the interaction of 
ultra-high energy cosmic rays with the cosmic microwave background radiation 
(Halzen et al. 1990) and the unresolved point sources in extragalactic 
astronomical objects (Sigl et al. 1994).
In these cases, this results in an isotropic flux of secondary photons.
Therefore, the measurements of the diffuse isotropic gamma-ray flux
might provide information on the ultra-high energy cosmic ray components.
Moreover, this flux would represent a background for experiments searching for
the gamma-ray enhancement from the direction of the galactic disk.

Up to energies well into the GeV region gamma rays are 
best observed from quite a number of satellites which have 
been active during the last decades (Ackermann et al. 2015). 
But with their limited detector size, typically well below $1\,$m$^2$,
they run out of statistics at higher energies, due to the steep decrease of
intensity of all types of primary cosmic ray particles.
Therefore, only ground based experiments are capable to
extend the observations to higher energies.

Gamma-ray astronomy in the upper GeV and TeV regions
has seen a dramatic development during the last decade. This
was possible by the development of the imaging atmospheric
Cherenkov (IAC) technique (Abramowski 2014).
Here, the primary particle is not observed directly but indirectly 
via the large number of secondary particles produced by its 
interactions in the Earth's atmosphere. 
Most of these are relativistic electrons which
emit Cherenkov light since they move faster than the speed of
light in air. It has been possible, by this technique, to identify
gamma rays from astronomical objects up to about 100 TeV. 

Extending these measurements to higher energies is again
hampered by poor statistics. IAC telescopes have an effective
acceptance area of several times $10^4\,$m$^2$. 
But their field of view is small, below $5^\circ$, and they 
can only take data in dark and clear nights which limits their observations 
to 10-15\% of real time even at the most suitable observation sites.
Non-imaging Cherenkov light measurements, like the Tunka-133 
experiment (Prosin et al. 2014) provide the coverage of larger areas, but are 
still hampered by the request to moonless clear nights.   
Therefore, air-shower detectors which measure the charged particles 
produced by the high-energy primary in the atmosphere have
been employed to search for cosmic gamma rays above 100 TeV. 
Their angular acceptance is well above $60^\circ$ and they can take
data around the clock.

Extensive air showers are mainly characterized by the total electron number 
and the total muon number.
In general, muons are produced by the decay of charged kaons or pions,
which in hadronic showers are produced in nucleus-nucleus interactions,
whereas in photon generated showers only by photoproduction processes.
The ratio between the cross sections of photoproduction and 
nucleus-nucleus interaction processes is very small, in the order of 
$\sim 10^{-3}$.
Therefore, the usual strategy for searches for primary gamma rays 
in extensive air showers
is to discriminate gamma-ray primaries from the hadronic background
by identifying muon-poor or even muon-less extensive air showers.

This article presents upper limits on the relative intensity of
the gamma-ray component of cosmic rays
from the measurements by the KASCADE and KASCADE-Grande experiments.
In particular, the results of the original KASCADE experiment are updated by 
including eight years more data since the publication in 2003 (Schatz et al.).
For the present results of KASCADE-Grande, nearly five times more data sets of the 
gamma-ray simulations are used, compared to the one for the initial result in 
(Kang et al. 2015), whereas the same method of the analysis is applied.

Earlier measurements of this kind have claimed that about
1 in 1000 cosmic ray particles are gamma rays
at energies around PeV
(Nikolsky et al. 1987). More recent experiments (Chantell et al. 1997;
Aglietta et al. 1996) were unable to confirm these results
and yielded only upper limits of order $10^{-4}$ for the gamma-ray fraction
at energies from $5.7\times10^{14}$ eV to $5.5\times10^{16}$ eV.
Recently, a re-analysis of the Moscow State data (Kalmykov et al. 2013) claimed the 
observation of detected gamma rays at an energy of around 100 PeV.
A claim which is still under discussion (Fomin et al. 2017).
We have therefore considered it worthwhile to
search the full measurements taken by the cosmic ray experiments
KASCADE and its extension KASCADE-Grande for events induced by primary gamma rays.
For the present analysis we have chosen the traditional method following 
Helene (1983) in order to compare our results directly with results of earlier studies.

\section{Experiments and data selection}
\subsection{The KASCADE experiment}
The KArlsruhe Shower Core and Array DEtector (KASCADE) 
was located at the Karlsruhe Institute of Technology,
Germany ($8.4^{\circ}$E, $49.1^{\circ}$N) at 110 m above sea level, corresponding
to an average vertical atmospheric depth of 1022 g cm$^{-2}$. 
It operated until the end of 2012 and all components are meanwhile dismantled.
The experiment measured the electromagnetic, muonic and hadronic
components of extensive air showers with three major
detector systems: a large field array, a muon tracking detector system
and a central detector. A detailed description of the KASCADE
experiment can be found in Ref. (Antoni et al. 2003).
\begin{figure}[t!]
\begin{center}
\includegraphics[width=0.45\textwidth]{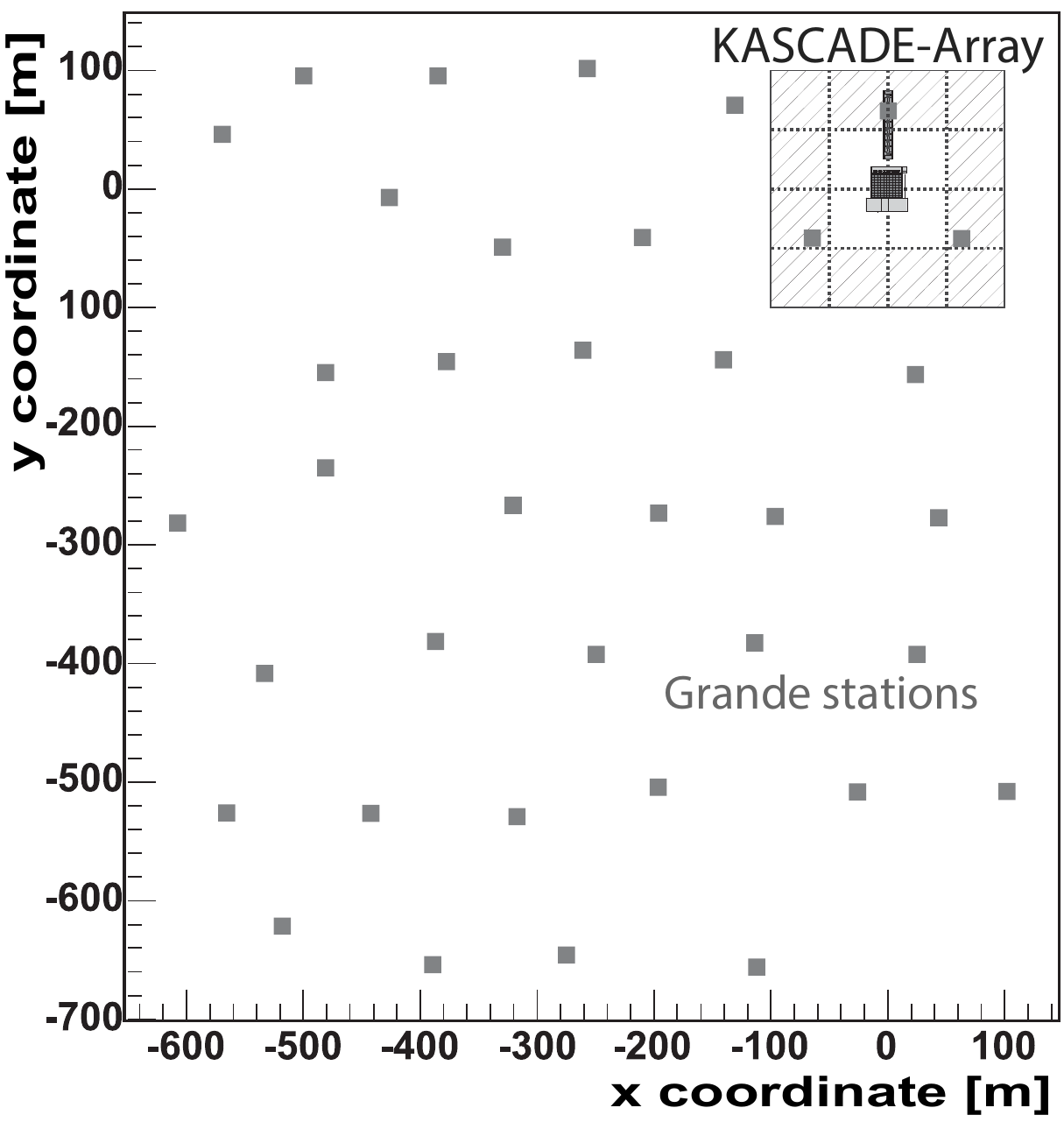}
\caption{
  Schematic view of the KASCADE and KASCADE-Grande experiments.
  The shaded area is covered by 622 m$^{2}$ of muon detectors.
  The KASCADE central detector and muon tunnel are also shown, but not used in this analysis.}
\label{fig1}
\end{center}
\end{figure}

In this analysis, data from the 200$\times$200 m$^{2}$
scintillation detector array are used. The 252 detector stations
are uniformly spaced on a square grid of 13 m separation. The
stations are organized in 16 electronically independent
clusters with 16 stations in the 12 outer clusters and 15
stations in the 4 inner clusters (see Fig.~\ref{fig1}). 
The stations in the inner and
outer clusters contain four and two liquid scintillation detectors,
respectively, covering a total area of 490 m$^{2}$. In addition,
plastic scintillators are mounted below an absorber of 10 cm
of lead and 4 cm of iron in the 192 stations of the outer
clusters (622 m$^{2}$ total area). The absorber corresponds to 20
electromagnetic radiation lengths, representing a threshold for
vertical muons of 230 MeV. This configuration allows the
measurement of the electromagnetic and muonic components
of extensive air showers. 
\begin{figure}[t!]
\begin{center}
\includegraphics[width=0.45\textwidth]{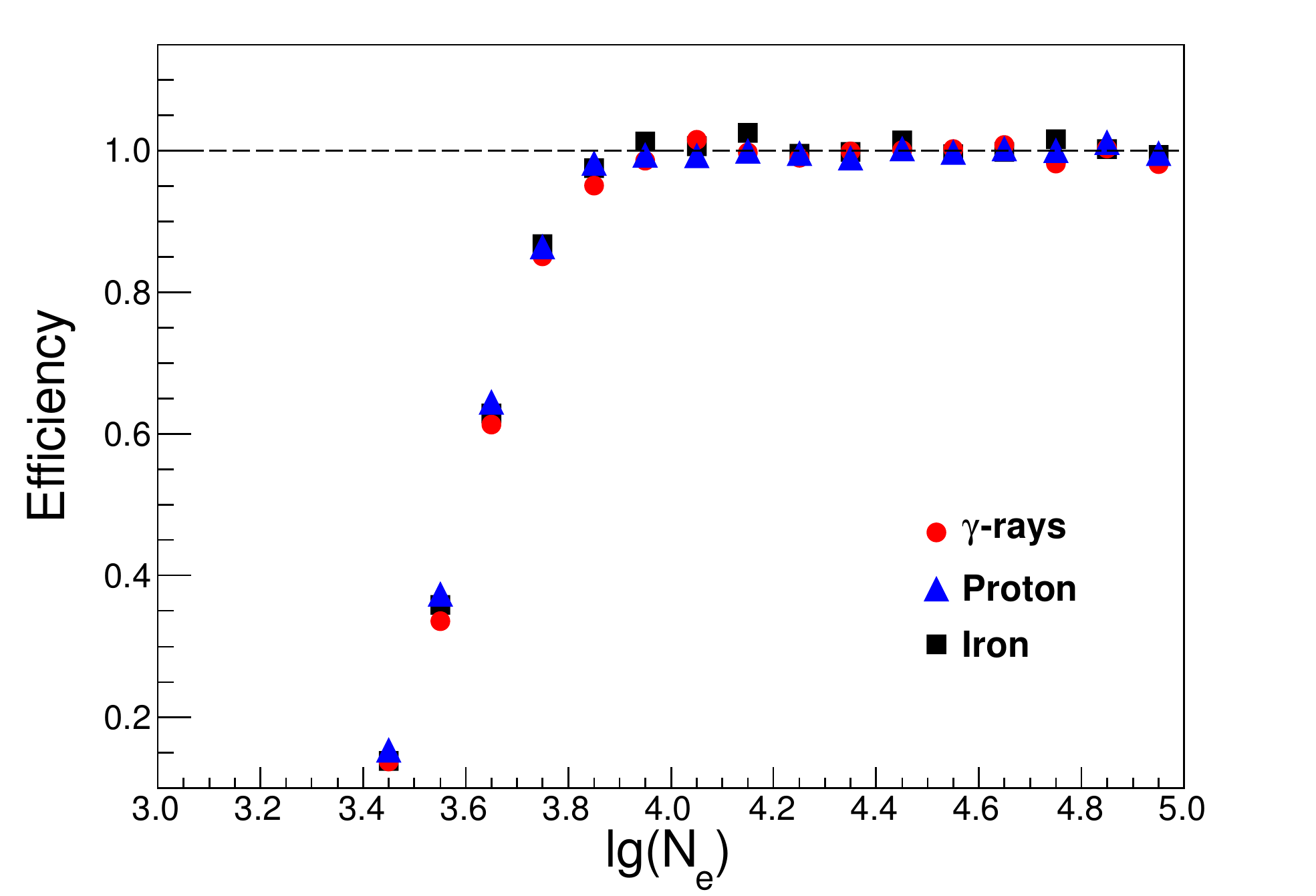}
\includegraphics[width=0.44\textwidth]{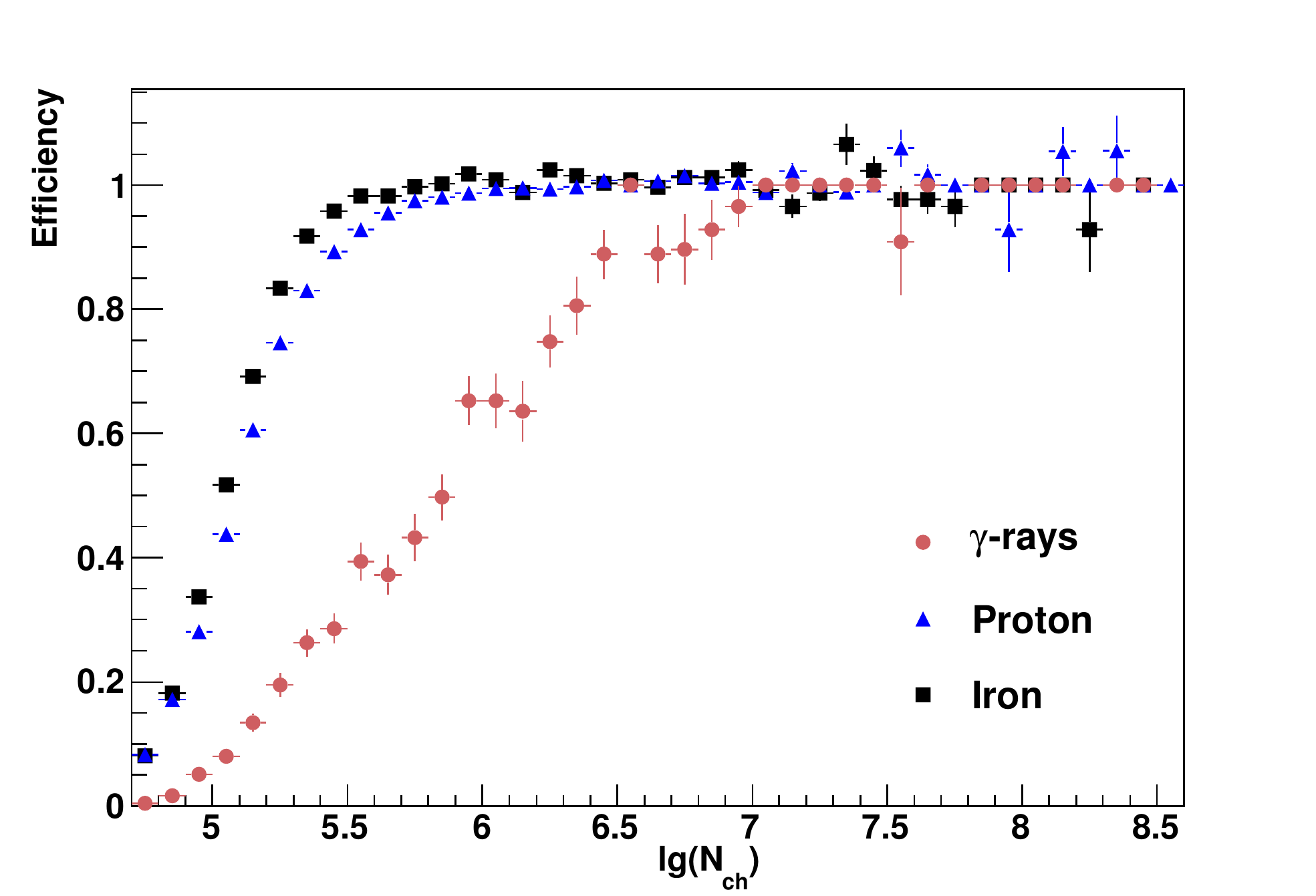}
\caption{
Trigger and reconstruction efficiency as a function of the number of electrons for KASCADE (top) 
and the number of charged particles for KASCADE-Grande (bottom)
for air showers induced by photons, protons and iron primaries.}
\label{fig2}
\end{center}
\end{figure}

Applying an iterative shower reconstruction procedure, the
numbers of electrons and muons\footnote{In fact, in KASCADE the so named truncated muon number is 
used, which integrates the muon lateral distribution from $40-200\,$m only in order to reduce the 
effects of electromagnetic punch-through at low distances and the uncertain shape of the lateral 
distributions at large distances.}   
in a shower are determined
basically by maximizing a likelihood function describing the
measurements with the Nishimura-Kamata-Greisen (NKG)
formula (Greisen 1956; Kamata \& Nishimura 1958),
assuming a Moliere radius of 89 m. 
Detector signals are corrected for contributions of 
other particles, e.g., the signal of the electromagnetic detectors
for contributions from muons, secondary $\gamma$ rays, and hadrons (Antoni et al. 2001). 
The directions of the incoming primary particles 
are determined without assuming a
fixed geometrical shape of the shower front by evaluating the
arrival times of the first particle in each detector and the total
particle number per station.

In this analysis, to remove from the data set poorly reconstructed showers 
and to obtain the precise muon number measurement, 
the following cuts were applied: all the 16
clusters should be working, shower core positions have to lay inside
a circular area of 91 m radius around the center of the KASCADE array 
to avoid large reconstruction errors at the edges of the
detector field and zenith angles are required to be $\theta<20^{\circ}$,
where the angular resolution of KASCADE is better than $0.55^{\circ}$ 
for electron numbers lg($N_{e}$) $>$ 4.
The data set was recorded between 1996 October and 2010 May,
which is corresponding to an effective time of about 4223.6 days. 
About 1.0$\times10^{8}$ events are remaining for the analysis 
after the cuts.

\subsection{The KASCADE-Grande experiment}
The KASCADE-Grande (Apel et al. 2010) array covering an area of 700$\times$700\ m$^{2}$ was
an extension of the original KASCADE array and was
optimized to measure extensive air showers up to primary energies of 1\ EeV (Fig.~\ref{fig1}).
It comprised 37 scintillation detector stations located on a
hexagonal grid with an average spacing of 137\ m for the measurements of the
electromagnetic and muonic shower components.
Each of the detector stations was equipped with
plastic scintillators covering a total area of 10\ m$^{2}$.
The muon information is taken from the KASCADE muon detectors.

For this analysis, full data sets taken from 2003 to 2012 were used,
where only successfully reconstructed and precisely measured events were selected.
The core positions of the showers are inside
an area of 152,214 m$^{2}$ 
around the center of the Grande array due to the same reason as KASCADE.
The zenith angle must be smaller than 40$^{\circ}$ to ensure full efficiency and 
an angular resolution better than 0.5 - 0.8$^{\circ}$ 
over the whole energy range.
After applying all quality cuts,
we obtained in total ca. $1.7\times10^{7}$ events for a measurement time of ca. 1865 days.

\section{Gamma and cosmic ray simulations}
An essential part of the present analysis is 
the Monte-Carlo simulation of the shower development and the 
response of the experiment.
Showers initiated by primary photons, protons, He, CNO, Si and iron nuclei have been 
simulated in order to estimate the mean energies of gamma rays and cosmic rays, respectively.  
However, the optimisation of the selection of primary gamma candidates is
based on the comparison of measured events with simulations of primary gamma-rays, only.

For the simulation of the physical processes in the air shower development
the CORSIKA 6.9 (Heck et al. 1998) program has been used.
To determine the signals in the individual detectors, all secondary
particles at the ground level are passed through a complete detector simulation program
using the GEANT 3 (Brun et al. 1987) package.
The predicted observables at ground level, such as e.g. the number of
electrons, muons and photons are then compared to the measurements.
The FLUKA 2002.4 (Fasso et al. 2005) model
has been used for hadronic interactions at low primary energies (E $<$ 200\ GeV).
High-energy interactions were treated with a different model, in our case by 
QGSJET-II-02 (Ostapchenko 2006).
The same number of showers was generated for each particle type.
The simulations cover the energy range of $10^{14} - 10^{18}$ eV
and the zenith angle interval $0 - 60^{\circ}$.
The zenith angle distribution follows that of an isotropic flux, where the mean values after
considering threshold effects are 12.7$^{\circ}$ and 20$^{\circ}$ for KASCADE and KASCADE-Grande, 
respectively. The azimuth distribution is flat and uniform over 360$^{\circ}$ for both arrays.
The simulations were performed with an $E^{-2}$ spectral index for photons and cosmic rays as well.
In total, ca. 2.5 million events are generated for cosmic rays and also for gamma rays.

\subsection{Estimation of cosmic ray energy}
The energy of cosmic rays is estimated from the experimental data.
For KASCADE, the formula based on the measurement of the muon and electron numbers 
of the KASCADE array (Glasstetter et al. 2005) is used. The uncertainties of
the energy reconstruction have been studied with Monte-Carlo simulations.
For events of this selection, the typical uncertainty is
about 30\% for individual events,
caused mainly by the unknown cosmic-ray composition.

For KASCADE-Grande, the cosmic ray energy is derived from independent measurements 
of the charged particles and muon components of the secondary particles of extensive air showers
by the KASCADE-Grande experiment (Apel et al. 2011),
where all muon identification comes from KASCADE.
The measured data are analyzed on a single-event basis
taking into account the correlation of the two observables.
The resulting systematic uncertainty
in the flux of the cosmic rays
is estimated to be of 10 - 15\%, 
based on the hadronic interaction model QGSJET-II-02
\footnote{      
It is worth to mention that there is a difference in the shower muon content of extensive air showers due to 
effects of the modified treatment of charge exchange processes in pion collisions between the QGSJET-II-02 and 
the most recent QGSJET-II-04 models (Kang et al. 2013). 
However, the change of the final results for the present analysis is expected to be negligible as we compare 
gamma-ray simulations with measured data for the selection of gamma-ray candidate events.}.
\begin{figure}[h]
\begin{center}
\includegraphics[width=0.43\textwidth]{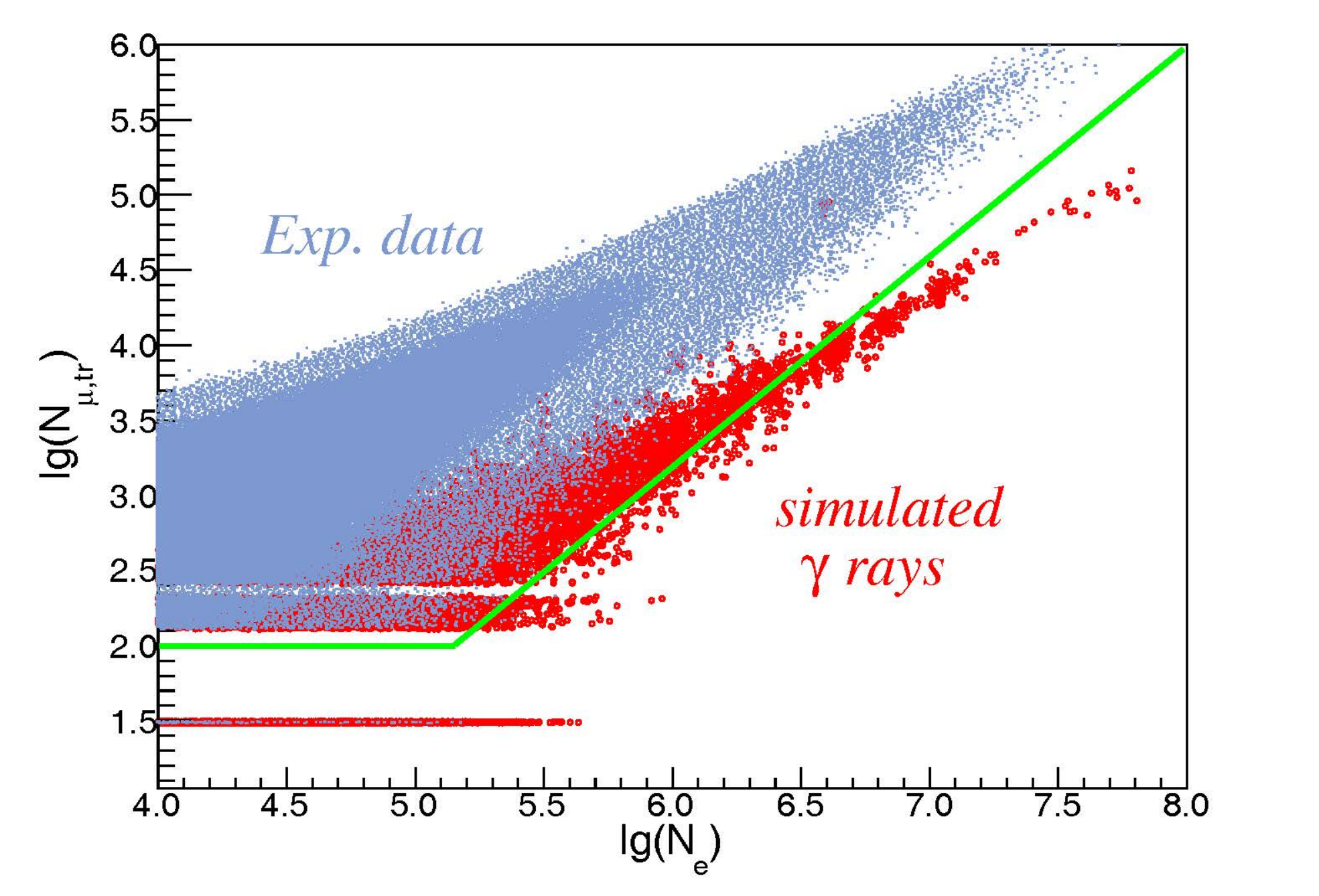}
\includegraphics[width=0.43\textwidth]{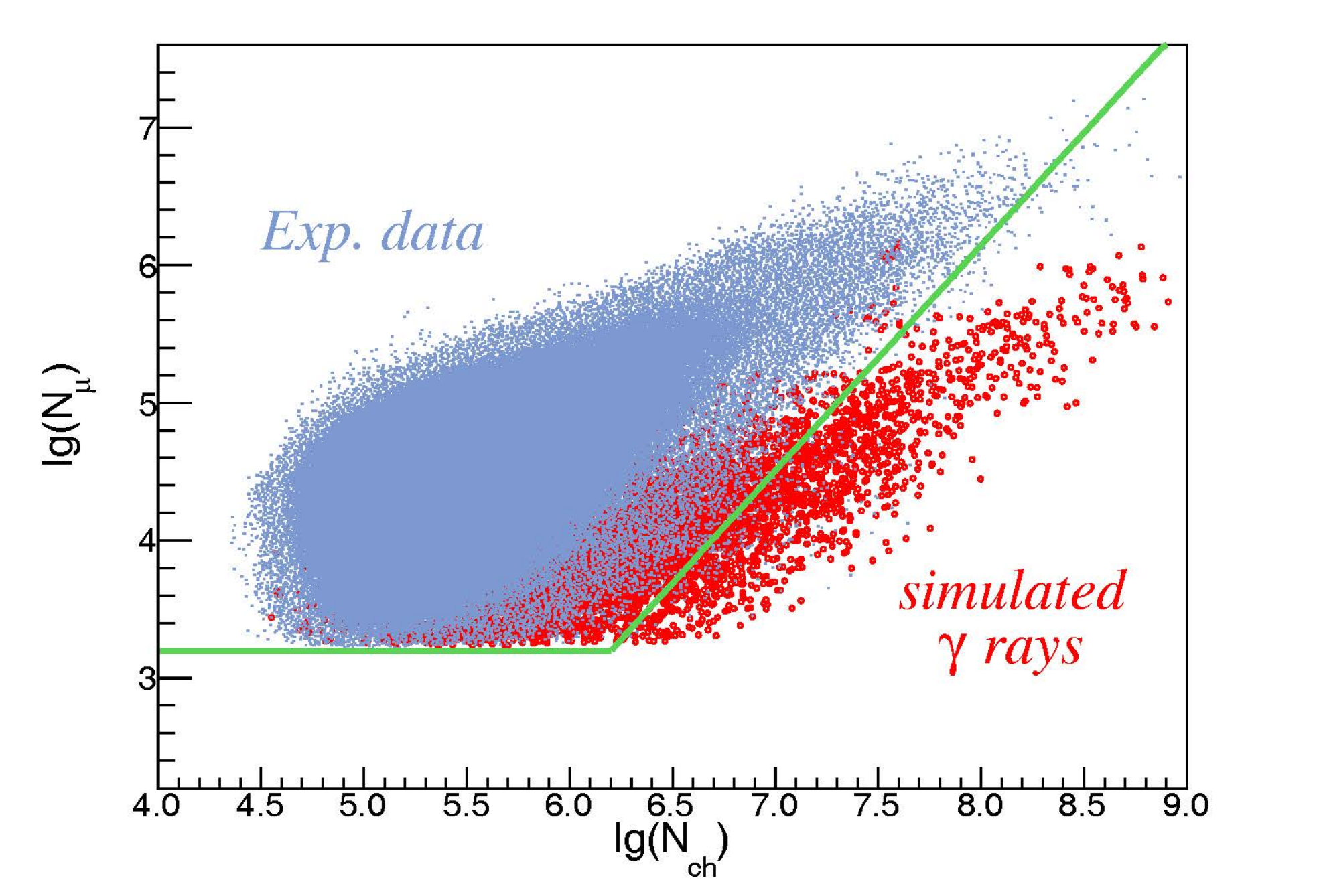}
\caption{
Scatter plot of the measured number of muons lg$(N_{\mu,tr.})$
vs. number of electrons lg$(N_{e})$ for KASCADE (top)
and for number of muons lg$(N_{\mu})$ vs. number of charged particles lg$(N_{ch})$ for 
KASCADE-Grande (bottom).
In both cases simulated gamma-ray showers are superimposed. 
The lines indicate the selection criteria for the subset of the muon-poor showers.}
\label{fig3}
\end{center}
\end{figure}

\subsection{Efficiency}
For calculating the upper limits of the gamma-ray component of the cosmic-ray flux the efficiencies of the 
experiment has to be known.
The trigger and reconstruction efficiency of both experiments is shown in Fig.~\ref{fig2}.
The efficiency is defined by the ratio of the number of reconstructed events 
in the area considered to the number of simulated events falling in this area, 
where binomial statistical errors are used. 
Efficiency above one is possible due to the fact that by the core uncertainty eventually more 
events are reconstructed inside the area than simulated.

For the KASCADE experiment, the trigger and reconstruction efficiency as a function of the shower size,
i.e. the number of electrons, is demonstrated in the upper graph of Fig.~\ref{fig2}. 
The detector array reaches full efficiency on the detection of showers for electron numbers
lg($N_{e}$) $>$ 4 for air showers induced by $\gamma$ rays, protons, and iron primary
particles, which approximately corresponds to a primary energy of 
$2.5\times10^{14}$ eV for $\gamma$ rays and $3.3\times10^{14}$ eV for charged cosmic rays.

The trigger and reconstruction efficiency of KASCADE-Grande as a
function of the shower size, i.e. the number of charged particles, is
demonstrated in the lower graph of Fig.~\ref{fig2}.
Full efficiency is reached at the number of charged particles of around
10$^{6}$ for air showers induced by protons and iron primary particles,
which corresponds to a primary energy of about 10$^{16}$\ eV.
However,
for showers induced by photon primaries, full efficiency
is reached at a higher number of charged particles due to the missing muon 
trigger at large distances.
The limit at high energies is due to the restricted area of the Grande array.

\section{The analysis}

\subsection{Gamma-hadron discrimination}
Since gamma-ray induced air showers are notable for their lack of muons compared to hadronic showers,
we select a data sample enriched in gamma showers by rejecting showers containing muons.
Simulations of hadronic air-showers underlie large systematic uncertainties
due to be above the interaction energies of accelerators and the unknown elemental composition.
Therefore, our selection concept is based on using the comparison of the measured events
with simulated gamma-ray induced showers.

The upper graph of Fig.\ \ref{fig3} shows the distribution 
of the electron versus truncated muon number lg($N_{\mu,tr})$ - lg($N_{e}$) 
for measured showers by KASCADE with simulated $\gamma$-ray-induced showers.
The graph indicates the whole experimental data set, 
as well as the distribution of the simulated $\gamma$-ray events.
Here $N_{\mu,tr}$ denotes the number of muons 
in the distance range of 40 - 200 m from the shower core.
The electron number $N_{e}$ is corrected to a zenith angle of $12.7^{\circ}$ 
using an attenuation length of $\Lambda_{N_{e}}$ = 158 g cm$^{-2}$ (Antoni et al. 2003).
In Fig.\ \ref{fig3}, the showers without any detected muons
are plotted with lg$(N_{\mu,tr})$ = 1.5 to be visible at the logarithmic axis
\footnote{
In KASCADE, it is also possible to detect only one, two, or three
muons per shower. When using these low numbers of measured
muons to calculate the total muon numbers, almost fixed
values are obtained, smeared only by the variation of the core
location and the zenith angle of the individual showers
(see Fig.\ \ref{fig3}).
}.
The distribution of the shower size for $\gamma$-ray-induced showers 
motivates the following cuts to select the muon-poor showers: 
lg($N_{\mu,tr}) < 2$ for lg($N_{e}) < 5.15$ and lg($N_{\mu,tr}) 
< 1.4\cdot$lg($N_{e})-5.21$ for lg($N_{e}) > 5.15$.
This selection of the muon-poor showers, i.e. $\gamma$-ray candidates,
is indicated by straight lines in Fig.\ \ref{fig3} (top).

For KASCADE-Grande, the distribution of the observed number of muons lg($N_{\mu}$)
versus the charged particles lg($N_{ch}$) is shown in the lower graph of Fig.\ \ref{fig3},
where for KASCADE-Grande never zero muons were detected
\footnote{For KASCADE-Grande showers the muons are measured at larger distances to the shower core at high energies (KASCADE: 40 - 200 m). In all showers (simulated and measured, $\gamma$-ray or proton induced) at least one muon was detected.}.
The simulated gamma-ray showers are superimposed as well.
The attenuation of the shower size through the atmosphere
is corrected to a zenith angle of 20$^{\circ}$
using the method of constant intensity cut (Apel et al. 2012).
The selection of muon-poor showers is indicated by the straight line in Fig.\ \ref{fig3} (bottom):
lg($N_{\mu}) < 3.2$ for lg($N_{ch}) < 6.2$ and lg($N_{\mu}) < 1.64\cdot$lg($N_{ch})-6.95$ for lg($N_{ch}) > 6.2$.

The stringent selection cuts are motivated by the simulated gamma-rays
and its optimization was investigated by changing the slope of the selection line.
The cut values were found and optimized by Monte-Carlo simulations in order to maximize
the purity/efficiency ratio of $\gamma$-ray induced events.
The events below the straight line were taken into consideration for further analysis.
They amount to 1056 out of a total of 17 millions events in case of KASCADE-Grande
and to 12087 out of 100 millions in case of KASCADE.
In the region below this line the events are expected to be mainly due to primary gammas
because air showers induced by heavy nuclei show a larger muon to electron ratio.
By this, we therefore use a conservative way in searching for gamma-ray induced showers 
in which the expected background is not subtracted from the event number below the 
selection lines shown in Fig.\ \ref{fig3}.
\begin{figure*}[t!]
\begin{center}
\includegraphics[width=0.65\textwidth]{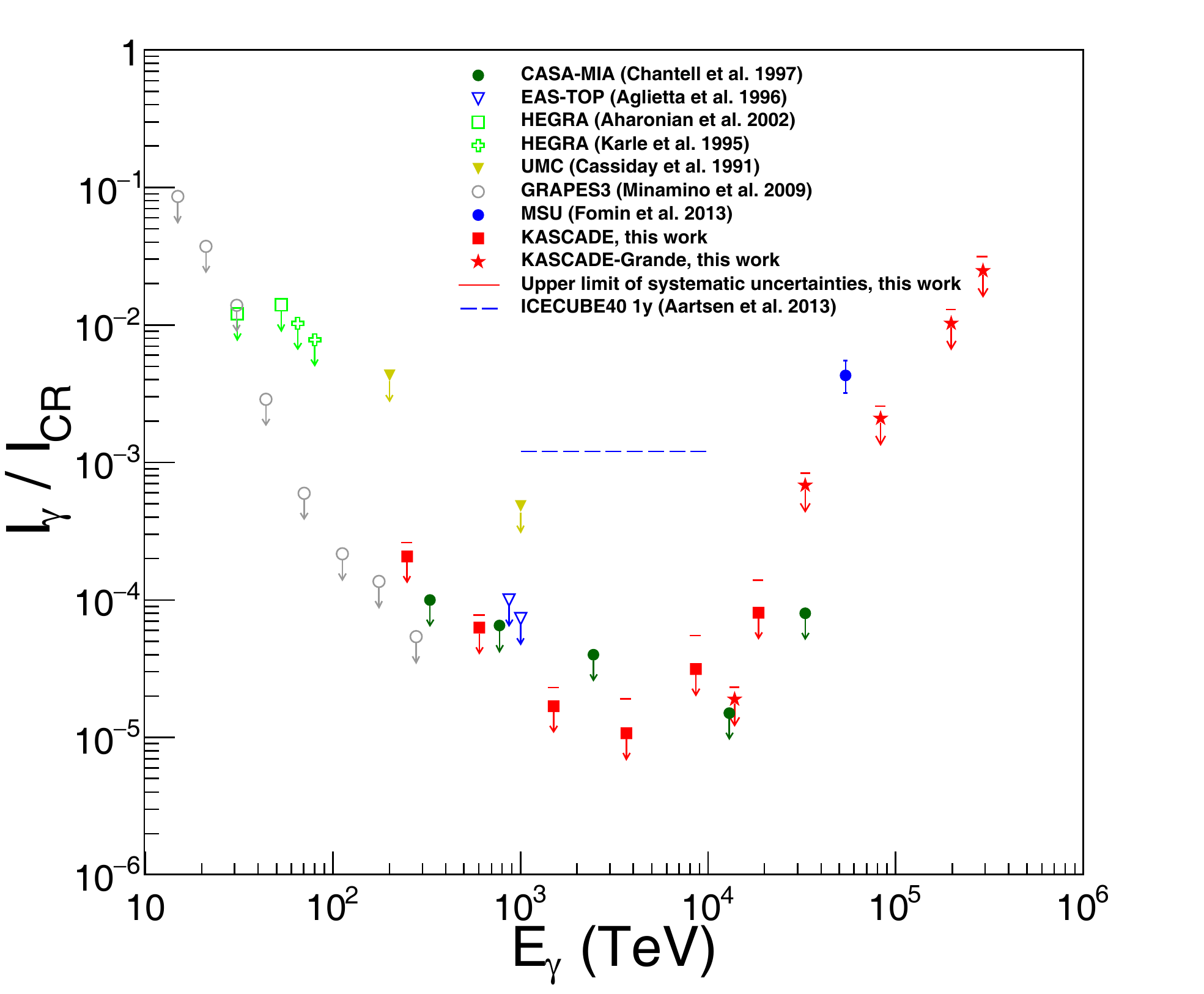}
\caption{
Measurements of the fraction of $\gamma$ rays relative to
cosmic rays in the energy from $10^{13}$ eV to $10^{18}$ eV.
The points with arrows represent
upper limits from the CASA-MIA (90\% C.L., Chantell et al. 1997),
EAS-TOP (90\% C.L., Aglietta et al. 1996), 
HEGRA (90\% C.L., Karle et al. 1995; Aharonian et al. 2002),
UMC (90\% C.L., Matthews et al. 1991), 
GRAPES3 (90\% C.L., Minamino et al. 2009),
IceCube (90\% C.L., Aartsen et al. 2013),
except the MSU (95\% C.L., Fomin et al. 2013) experimental value.
The red squares and stars represent the results
from KASCADE (90\% C.L.) and KASCADE-Grande (90\% C.L.), respectively,
with systematic uncertainties.
Limits reported by the Tibet array (3-10 TeV, 90\% C.L., Amenomori et al. 2002) 
and by Milagro (3.5-15 TeV, 90\% C.L., Abdo et al. 2008) are out of this energy range.
}
\label{fig4}
\end{center}
\end{figure*}
\begin{figure*}[t!]
\begin{center}
\includegraphics[width=0.7\textwidth]{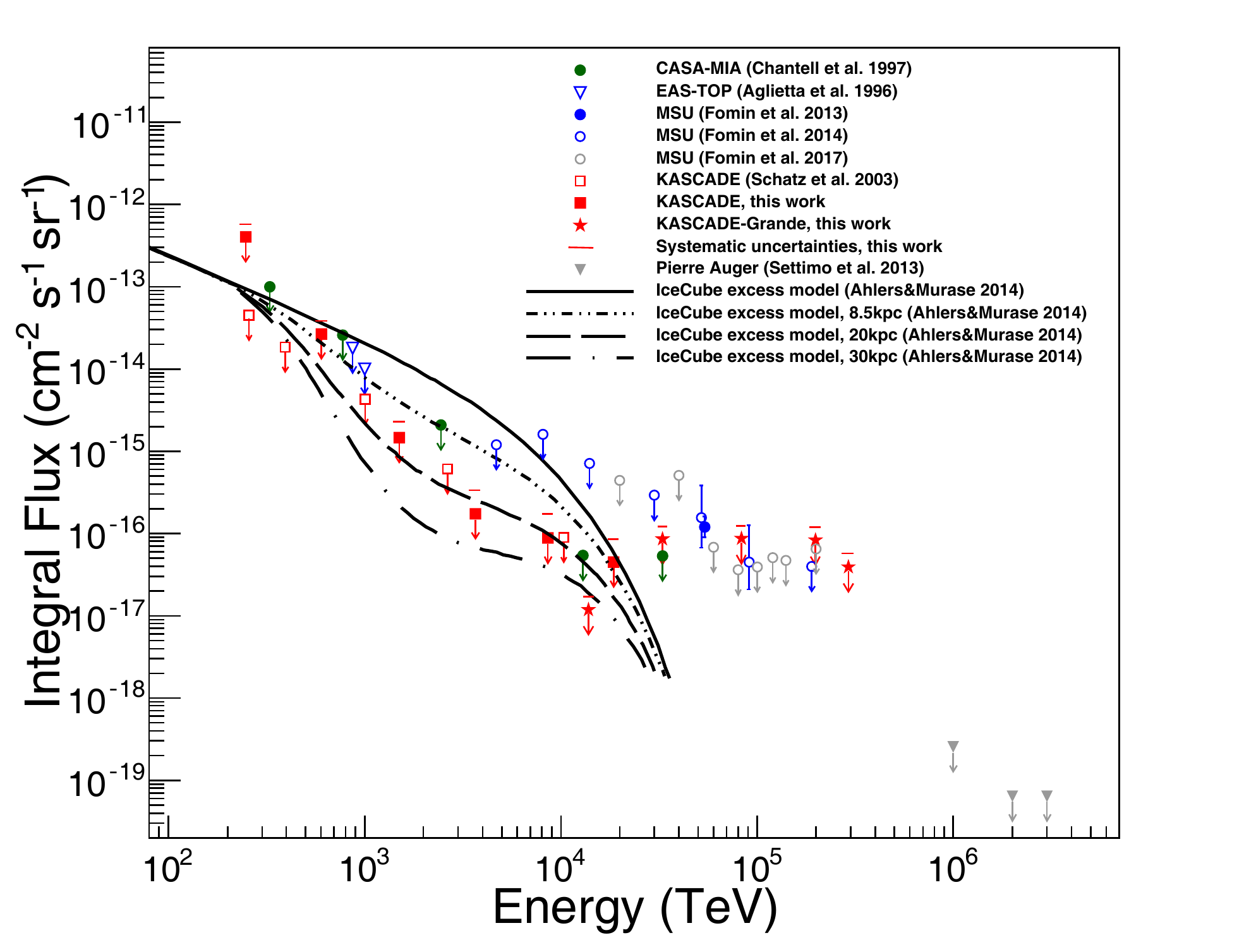}
\caption{Comparison of integral flux of gamma rays (including systematic uncertainties) with 
previous results and with theoretical curves by an IceCube excess model (Ahlers \& Murase 2014).
The lines are for the unattenuated flux (solid) and the flux from 
8.5 kpc (distance of the galactic center), 20 kpc, and 30 kpc, respectively.
The high-energy points from the Pierre Auger Observatory are taken from Settimo et al. (2013).
}
\label{fig5}
\end{center}
\end{figure*}

\subsection{Upper limit of $I_{\gamma}/I_{CR}$}
There is no excess of events consistent with a gamma-ray signal seen in the data. Hence,
we assume that all events below the selection line are primary $\gamma$-rays and set 
upper limits on the gamma-ray fraction of the cosmic rays.

To determine an upper limit on the fraction of the $\gamma$ ray with respect to the cosmic ray integral flux, $I_{\gamma}/I_{CR}$,
we use the equation given by Chantell et al. (1997):
\begin{equation}
\frac{I_{\gamma}}{I_{CR}}<\frac{N_{90}}{N_{tot}\epsilon_{\gamma}}\left(\frac{E_{CR}}{E_{\gamma}}\right)^{-\beta+1}.
\end{equation}
$N_{tot}$ is the total number of events and
$N_{90}$ is the 90\% C.L. upper limit on the number of detected events, 
which is estimated by means of the statistical method from Helene (1983).
Using the gamma simulations, we estimate the efficiency for $\gamma$-ray detection and 
reconstruction, $\epsilon_{\gamma}$, after applying the selection cuts.
$E_{CR}$ and $E_{\gamma}$ are the mean energies of cosmic rays and gamma rays, respectively, 
which produce the same shower size.
$\beta$ is the spectral index of the integral flux of cosmic rays:
$\beta$ = 2.7 for $E<4\times10^{15}$eV and $\beta$ = 3.0 for $E>4\times10^{15}$eV.
That means, to calculate the energy we weight the simulated energy spectra of $\gamma$-ray and 
cosmic-ray showers to an $E^{-2.7}$ below knee and $E^{-3}$ above knee, 
assuming the knee is located at $4\times10^{15}$ eV, as indicated by the measurements of Antoni et al. (2005).
The mean energies of cosmic rays and gamma rays are then calculated bin by bin.
$\gamma$-ray primaries produce larger showers than cosmic-ray primaries at the same energy.
For this reason, to convert fixed shower size to fixed energy of gamma-ray primary,
different factors  in the equation has to be taken into account
with the assumption that both components have the same spectral index.

Contributions to the systematic uncertainties are mainly due to the quality of discrimination 
of $\gamma$ rays from cosmic rays (i.e., the statistics of  the $\gamma$-ray simulations), 
the estimation of the cosmic-ray energy (i.e., the weighting of the energy spectral index and 
unknown composition), and the uncertainty in the validity of the hadronic interaction models.

\section{Results and discussion}
The results of the search for diffuse ultra-high energy gamma rays for different threshold values 
of $N_{ch}$ are summarized in Table~\ref{tb1} and Table~\ref{tb2} for both experiments.
Detailed systematic uncertainties on the upper limit of $I_{\gamma}/I_{CR}$ are 
determined and also listed in the tables. We considered here the uncertainties on the energy 
reconstruction of the cosmic rays and gamma rays as well as on the slope of the cosmic ray 
spectrum and the efficiency of the cut.
The uncertainties of the estimation of the cosmic-ray and gamma-ray energies are estimated to 
be about 20\% (Antoni et al. 2005; Apel et al. 2011) and 10\%, respectively.  
Other sources of systematic uncertainties, like variation of hadronic interaction models or 
the unknown chemical composition as well as the uncertainties in the slope of the all-particle 
energy spectrum is included in these numbers.
The error on the efficiency for the gamma-ray detection is evaluated to be smaller than 20\%. 
Thus, the resulting uncertainty is less than 25\% in the flux limits, except the three highest 
energy bins of KASCADE, since there the number of events at 90\% C.L. limit is estimated from 
zero observed events.

To determine upper limits to the integral flux of gamma rays at fixed gamma-ray energies,
we use measurements of the all-particle primary energy spectra reported in references 
(Antoni et al. 2005; Apel et al. 2011).
The limits on the gamma-ray flux $I_{\gamma}$ are also listed in 
Tables ~\ref{tb1} and ~\ref{tb2}.
To obtain the integral flux of gamma rays, we multiplied the reconstructed all-particle energy spectrum 
by the fraction of gamma rays relative to cosmic rays $(I_{\gamma}/I_{CR})$.

Figure \ref{fig4} displays the measurements on the gamma-ray fraction as a function of the energy, 
including this work, for the energy range of $10^{14}$~eV up to $10^{18}$~eV.
The upper limit of the fraction of $\gamma$-rays at $1.5\times10^{15}$~eV and $3.7\times10^{15}$~eV
are obtained to be $1.7\times10^{-5}$ and $1.1\times10^{-5}$, respectively.
These are the lowest upper limits up to now. 
In addition, as around $10^{17}$~eV (Fig.\ \ref{fig4}) not many experiments have reported results,
the limits obtained by KASCADE-Grande are of heightened interest. 
It should be noted that all values in Fig.\ \ref{fig4} are upper limits,
except the one from MSU (Fomin et al. 2013, Fomin et al. 2014).
This positive signal, however, is in conflict with the limits presented here. Also, a 
further reanalysis of the MSU data does not confirm the positive signal (Fomin et al. 2017).

Figure \ref{fig5} shows the comparison of integral flux of $\gamma$ rays 
with other previous experiments (Aglietta et al. 1996; Chantell et al. 1997).
Compared with the earlier obtained limits by KASCADE in 2003 (Schatz et al. 2003) 
there are some differences, in particular, at lower energies.
This is accounted for by a more detailed investigation using more parameters 
(i.e., the lateral slope and the smoothness of the electromagnetic component), which affected 
mainly the lower energies. In this work we focused on higher energies by applying the same 
method to KASCADE and KASCADE-Grande data, where those additional parameters for selecting 
muon poor events are not effective.
The MSU collaboration (Fomin et al. 2014; Fomin et al. 2017) presented recently upper limits 
on the diffuse $\gamma$-ray flux for primary energies around 10-100 PeV obtained from an 
re-analysis of their old measurements. They are compatible with the presented KASCADE-Grande 
limits.

Figure \ref{fig5} also compares the results
exemplary with theoretical curves by a specific IceCube neutrino excess model
assuming proton-proton($pp$)-interactions and cutoff at 6 PeV (Ahlers \& Murase 2014).
The lines are the IceCube excess models
originating from different distances of sources of neutrinos in the galaxy,
where these neutrinos are also responsible for primary gamma-rays.
The secondary pions of the hadronic interactions of cosmic rays decay at the source
and produce a flux of high-energy neutrinos as well as gamma rays.
The relative number of neutrinos and gamma rays depend on the ratio of charged to neutral pions.
The flux limits on the gamma-ray flux $I_{\gamma}$ of this work
at $1.5\times10^{15}$ eV and $3.7\times10^{15}$ eV 
are lower than the theoretical prediction of the IceCube excess model 
coming from a 8.5 kpc source distance (dotted black line), which corresponds to the distance 
of the galactic center. Therefore, this observation is not in contradiction to the statement 
that the IceCube excess is associated with extragalactic sources, e.g. $\gamma$-ray bursts or 
active galactic nuclei.
Moreover, there are also other scenarios, e.g.~the PeV dark matter (DM) decay model in the 
galactic halo, suggested for explaining the IceCube excess (Ahlers \& Murase 2014), 
which can be additionally constrained by the results of this work, in a way as done by 
Kalashev and Troitsky (2015).

\begin{table*}[t!]
\begin{center}
\begin{tabular}{lcccccccc}
\hline\hline
lg($N_{e}$)&$N_{tot}$&$N_{90}$&$\epsilon_{\gamma}$&$E_{CR}$&$E_{\gamma}$&$I_{\gamma}/I_{CR}$&$I_{\gamma}$\\ \hline
$>$4   &$1.02\times10^{8}$ &11653.5 &0.33 &333   &248   &$< (2.1 \pm 0.5) \times10^{-4}$ &$<4.02\times10^{-13}$\\
$>$4.5 &$2.19\times10^{7}$ &583.3   &0.27 &783   &605   &$< (6.3 \pm 1.5) \times10^{-5}$ &$<2.67\times10^{-14}$\\
$>$5   &$3.92\times10^{6}$ &19.0    &0.18 &1994  &1502  &$< (1.7 \pm 0.6) \times10^{-5}$ &$<1.46\times10^{-15}$\\
$>$5.5 &$5.76\times10^{5}$ &2.3     &0.20 &5247  &3673  &$< (1.1 \pm 0.8) \times10^{-5}$ &$<1.73\times10^{-16}$\\
$>$6   &$6.75\times10^{4}$ &2.3     &0.44 &14618 &8603  &$< (3.1 \pm 2.4) \times10^{-5}$ &$<8.77\times10^{-17}$\\
$>$6.5 &$6.66\times10^{3}$ &2.3     &0.94 &44952 &18610 &$< (8.1 \pm 5.8) \times10^{-5}$ &$<4.47\times10^{-17}$\\
\hline\hline
\end{tabular}
\caption{\label{tb1}
Results of the search for diffuse ultra-high energy $\gamma$ rays 
at different threshold values of lg($N_{e}$) using KASCADE data. 
The median cosmic-ray energy, $E_{CR}$, and the median $\gamma$-ray energy, $E_{\gamma}$, 
are given in the fifth and sixth columns, respectively, in units of TeV. 
The quantities $N_{90}$ and $\epsilon_{\gamma}$ are defined in the text.
$I_{\gamma}/I_{CR}$ is the 90\% C.L. upper limit on the integral $\gamma$-ray fraction 
with systematic uncertainties,
and $I_{\gamma}$ is the 90\% C.L. upper limit on the integral $\gamma$-ray flux, 
in units of photons cm$^{-2}$s$^{-1}$sr$^{-1}$.}
\vspace*{0.5cm}
\begin{tabular}{llcccccc} \hline\hline
lg($N_{ch}$) & $N_{tot}$ & $N_{90}$ & $\epsilon_{\gamma}$ & $E_{CR}$ & $E_{\gamma}$ & $I_{\gamma}/I_{CR}$ &  $I_{\gamma}$ ($\times10^{-17}$)\\ \hline
$>$6.5 &6.19$\times10^{6}$ &358 &0.71      &3.21$\times10^{4}$ &1.38$\times10^{4}$ &$< (1.88 \pm 0.44) \times10^{-5}$ & $<$1.19\\
$>$7   &85537             &351 &$\sim$1.0 &8.72$\times10^{4}$ &3.29$\times10^{4}$ &$< (6.79 \pm 1.58) \times10^{-4}$ & $<$8.51\\
$>$7.5 &9640              &214 &$\sim$1.0 &2.21$\times10^{5}$ &8.30$\times10^{4}$ &$< (2.08 \pm 0.49) \times10^{-3}$ & $<$8.64\\
$>$8   &1239              &122 &$\sim$1.0 &5.31$\times10^{5}$ &1.98$\times10^{5}$ &$< (1.03 \pm 0.26) \times10^{-2}$ & $<$8.32\\
$>$8.5 &165               &78  &$\sim$1.0 &1.13$\times10^{5}$ &2.92$\times10^{5}$ &$< (2.48 \pm 0.68) \times10^{-2}$ & $<$3.89\\ \hline\hline
\end{tabular}
\caption{\label{tb2}
Results of the search for diffuse ultra-high energy gamma rays
at different threshold values of $N_{ch}$ using KASCADE-Grande data.
The mean cosmic-ray energy, $E_{CR}$, and the mean gamma-ray energy, $E_{\gamma}$,
are given in the unit of TeV.
$I_{\gamma}/I_{CR}$ is the 90\% C.L. upper limit on the integral gamma-ray fraction 
with systematic uncertainties,
and $I_{\gamma}$ is the 90\% C.L. upper limit on the integral gamma-ray flux
in the unit of photons cm$^{-2}$s$^{-1}$sr$^{-1}$. 
}
\end{center}
\end{table*}

\section{Conclusion and outlook}
Using data sets measured by the KASCADE and KASCADE-Grande experiment
over a period of 14 and 9 years, respectively,
we determined the 90\% C.L. upper limits to the diffuse flux of ultra-high energy gamma rays
for the energy range of $10^{14}$ eV to $10^{18}$ eV
by selecting showers with low muon content.

The upper limit of the fraction of $\gamma$ rays at $1.5\times10^{15}$ eV 
from the KASCADE measurement is estimated to be $1.7\times10^{-5}$, 
while  $1.1\times10^{-5}$ at $3.7\times10^{15}$ eV.
These are presently the lowest upper limits, 
which were used to set constrains on theoretical predictions, 
in particular, on the distance of sources 
for the IceCube neutrino excess model (Ahlers \& Murase 2014).

By means of the KASCADE-Grande measurements, 
the best upper limit to the fraction of the gamma-ray to the cosmic-ray flux
is obtained: $I_{\gamma} / I_{CR} < 1.88 \times 10^{-5}$ for 13.8 PeV.
The stringent limits above 100 PeV
might constrain a limit to the background rate of muon-poor showers
in the search for the galactic disk enhancement of cosmic rays.

The angular resolutions of KASCADE and KASCADE-Grande are sufficient over 
the whole energy range to search also for gamma-ray point sources, where a preliminary result was 
presented in Ref. (Kang et al. 2015a). 
However, both analyses will profit from an analysis of  combined KASCADE and KASCADE-Grande data, 
where a coherent shower reconstruction is presently under development.
By then also an advanced analysis method has to be optimized to give best results
(see, e.g. discussions in Homola and Risse 2007) for such a wide energy range
in the search for a diffuse gamma-ray flux.

\section*{Acknowledgments}
\acknowledgments
The authors would like to thank the members of the engineering and technical staff of the
KASCADE-Grande collaboration, who contributed to the success of the experiment. The
KASCADE-Grande experiment is supported in Germany by the BMBF and by the 'Helmholtz
Alliance for Astroparticle Physics - HAP' funded by the Initiative and Networking Fund of the
Helmholtz Association, by the MIUR and INAF of Italy, the Polish Ministry of Science and
Higher Education, and the Romanian Authority for Scientific Research UEFISCDI (PNII-IDEI
grants 271/2011 and 17/2011). J.C.A.V. acknowledges the partial support of CONACyT.
Z. Feng is supported by the Natural Sciences Foundation of China (Nos. 11405182)\\


\end{document}